\documentclass[manuscript]{aastex}
\usepackage{natbib}
\usepackage{graphicx,rotating,booktabs}

\newcommand{\mura}{$\mu_\alpha \cos{\delta}$}
\newcommand{\mudec}{$\mu_\delta$}

\newcommand{\mpy}{mas yr$^{-1}$}
\newcommand{\kmps}{km s$^{-1}$}

\slugcomment{To appear in ApJ.,}

\shorttitle{Proper motions in Taurus and Ophiuchus}
\shortauthors{Rivera et al.}

\begin{document}

\title{Internal and relative motions of the Taurus and Ophiuchus star-forming regions}

\author{Juana L. Rivera\altaffilmark{1}, 
Laurent Loinard\altaffilmark{1}, 
Sergio A. Dzib\altaffilmark{2}, 
Gisela N. Ortiz-Le\'on\altaffilmark{1},
Luis F.\ Rodr\'{i}guez\altaffilmark{1},
Rosa M. Torres\altaffilmark{3}}
\altaffiltext{1}{Centro de Radioastronom\'ia y Astrof\'isica, Universidad Nacional Aut\'onoma de M\'exico
Apartado Postal 3-72, 58090, Morelia, Michoac\'an, M\'exico}
\altaffiltext{2}{Max Planck Institut
f\"ur Radioastronomie, Auf dem H\"ugel 69, 53121 Bonn, Germany}
\altaffiltext{3}{Centro Universitario de Tonal\'{a}, Universidad de Guadalajara, Avenida Nuevo Perif\'{e}rico No. 555, 
Ejido San Jos\'{e} Tatepozco, CP 48525, Tonal\'{a}, Jalisco, Mexico}
\email{j.rivera@crya.unam.mx}

\begin{abstract}
We investigate the internal and relative motions of the Taurus and Ophiuchus star-forming regions using a sample of young stars with accurately measured radial velocities and proper motions. We find no evidence for expansion or contraction of the Taurus complex, but a clear indication for a global rotation, resulting in velocity gradients of order 0.1 km s$^{-1}$ pc$^{-1}$ across the region. In the case of Ophichus more data are needed to reliably establish its internal kinematics. Both Taurus and Ophiuchus have a bulk motion relative to the LSR (i.e.\ a non-zero mean peculiar velocity) of order 5 km s$^{-1}$. Interestingly, these velocities are roughly equal in magnitude, but nearly exactly opposite in direction. Moving back in time, we find that Taurus and Ophiuchus must have been very near each other 20 to 25 Myr ago. This suggests a common origin, possibly related to that of Gould's Belt. 

\end{abstract}


\section{Introduction}

Ophiuchus \cite[at 120 pc; ][]{Loinard2008} and Taurus \citep[at 130--160 pc;][]{Loinard2007,Torres2007,Torres2009,Torres2012} are two of the nearest star-forming regions \citep[see][for recent reviews]{Wilking2008,Kenyon2008}. They have both been instrumental in the emergence of our current understanding of low-mass star-formation \citep{Shu1987}, and have been studied extensively at virtually all wavelengths. For instance, there are extensive surveys of Taurus in X-rays by \cite{Guedel2007}, optical by Brice\~no et al.\ (1993; 1999), 
near-infrared by \cite{Duchene2004}, submillimeter by \cite{Andrews2005}, and radio by \cite{Dzib2015}. Similarly, in Ophiuchus, large-scale observations were obtained in X-rays \citep{Ozawa2005,Gagne2004}, near-infrared \citep{Haisch2002,Duchene2004}, submillimeter \citep{Motte1998,Johnstone2004}, and radio \citep{Dzib2013}. More recently, both regions have been targeted by the Spitzer Space Telescope \citep[e.g.,][]{Rebull2010,Padgett2008} and the Herschel Space Observatory (http://www.herschel.fr/cea/gouldbelt/en/). The distance to both regions is known very accurately thanks to recent trigonometric parallax measurements obtained from radio Very Long Baseline Interferometry (VLBI) observations \citep{Loinard2007, Loinard2008, Torres2007, Torres2009, Torres2012}. In the case of Taurus, the accuracy of these VLBI measurements is sufficient to characterize the depth of the complex, and crudely reconstruct its 3-dimensional structure \citep{Torres2007, Torres2009, Torres2012}.

The proximity of these two regions enables the detection of intrinsically faint sources (e.g.\ substellar objects) and ensures high linear spatial resolution. It also facilitates the accurate determination of proper motions, since for a given space velocity, the amplitude of the angular displacement diminishes linearly with distance. We will exploit this latter property here to derive the three-dimensional velocity vector for a sample of young stellar objects distributed across each of the regions. This will be achieved by combining radial velocity measurements from optical and near infrared spectroscopy, with proper motions derived from multi-epoch radio interferometric observations --supplemented, of course, by the accurate distances mentioned earlier to perform the conversion from angular to space velocity. Both conventional interferometers such as NRAO's Very Large Array (VLA) and long baseline interferometers (e.g.\ NRAO's Very Long Baseline Array; VLBA) can be used to obtain accurate proper motions. Examples of VLA measurements are shown in \cite{Loinard2003} and \cite{Chandler2005}, while examples of VLBA results can be found in \cite{Torres2007}.

In the present paper, we will collect existing radial velocity and radio proper motion measurements available in the literature for young stars in Taurus and Ophiuchus, and combine them to construct the 3-dimensional velocity vectors for roughly a dozen young stellar systems in Taurus and Ophiuchus. These results will be used to analyze the internal and relative kinematics of the Taurus and Ophiuchus star-forming regions.

\section{Compilation of radial velocities and proper motion measurements}

Two important technical  points must be made at the outset. The first one is that the proper motions measured using radio interferometers are, by construction, measured in a reference frame associated with the Solar System barycenter \cite[e.g.][]{Thompson2007}. The second is relative to the conversion of radial velocities from the LSR to heliocentric system. While the results of optical spectroscopy are usually reported in the heliocentric system, millimeter spectroscopic observations (that we will use for some of the sources) are often reported in the LSR system. For consistency with the proper motion measurements, we will express all radial velocities in the heliocentric system. The conversion from LSR to heliocentric involves the projection of the Sun motion along the line of sight. As we will see below, there is some on-going discussion about the true value of the Solar motion. However, as far as we know, all observatories use the same (fairly old) Solar motion determination, corresponding to +20 km s$^{-1}$ toward B1900 equatorial coordinates $(18^h; +30^\circ)$ for the conversion from heliocentric to LSR velocities. This is the value that we will use to perform the conversion when radial velocities in the literature are expressed in the LSR frame.

\subsection{Taurus}

In the case of the Taurus complex, accurate proper motion measurements are available for 7 young stellar systems: Hubble 4, HDE 283571, HP Tau G2, V773 Tau, T Tau, L 1551 IRS5, and DG Tau. We now briefly present and discuss each source invidually, emphasizing the distance, proper motion, and radial velocity measurements. A summary of these parameters is provided in Table 1.

\noindent {\bf Hubble 4} is a weak line T Tauri star of spectral type K7 located in the dark cloud Lynds 1495. Its trigonometric parallax ($\varpi$ = 7.53 $\pm$ 0.03 mas, corresponding to 132.8 $\pm$ 0.5 pc) and its proper motion (\mura\ = 4.30 $\pm$ 0.05 \mpy; \mudec\ = --28.9 $\pm$ 0.3 \mpy) have been measured using multi-epoch VLBA observations by \cite{Torres2007}. \cite{NGuyen2012} find evidence that Hubble 4 may be an SB2 spectroscopic binary, and provide multi-epoch measurements of the radial velocities of each of the two stars. The average value for the radial velocity of the primary is +18.0 km s$^{-1}$ with a dispersion of order 1.0 km s$^{-1}$ (all expressed in the heliocentric system). This is reasonably consistent with the older measurement of +15.0 $\pm$ 1.7 km s$^{-1}$ by \cite{Hartmann1986}. Conservatively, we will adopt +18.0 $\pm$ 2.0 km s$^{-1}$ for the radial velocity of Hubble 4.

\noindent {\bf  HDE 283572} (HIP 20388, V987 Tau) is a G2 star also located in the dark cloud Lynds 1495 with a VLBA parallax $\varpi$ = 7.78 $\pm$ 0.04 mas (128.5 $\pm$0.6 pc; Torres et al.\ 2007). Its proper motion, also measured with the VLBA, is \mura\ = 8.88 $\pm$ 0.06 \mpy; \mudec\ = --26.6 $\pm$ 0.1 \mpy\ \citep{Torres2007}. Its (heliocentric) radial velocity was measured by \cite{NGuyen2012} to be +14.2 $\pm$ 1.0 km s$^{-1}$. This is consistent with the value +15.0 $\pm$ 1.5 km s$^{-1}$ reported by \cite{Walter1988}.

\noindent {\bf HP Tau G2} (IRAS 04328-2248) belongs, together with the tight binary HP Tau G3, to a hierarchical triple system located on the eastern edge of the Taurus complex. Its proper motion, from VLBA observations, is \mura\ = 13.90 $\pm$ 0.06 \mpy; \mudec\ = --15.6 $\pm$ 0.3 \mpy\ \citep{Torres2009}, and its VLBA parallax is $\varpi$ = 6.20 $\pm$ 0.03 mas (161.2 $\pm$0.9 pc; Torres et al.\ 2009). Its radial velocity is +16.6 $\pm$ 1.7 km s$^{-1}$ according to \cite{NGuyen2012} and +17.7 $\pm$ 1.8 km s$^{-1}$ according to \cite{Walter1988}. We will adopt here the former of these two very consistent values.

\noindent {\bf V773 Tau} is a well-studied quadruple system in the Lynds 1495 cloud, at 132.8 $\pm$ 2.3 pc \citep{Torres2012}. The primary is a tight spectroscopic binary \citep{Welty1995} spatially resolved in VLBA observations \citep{Boden2007, Torres2012}. Its orbital motion has been very well characterized by \cite{Boden2007} and \cite{Torres2012} by combining spectroscopic and astrometric data. Two other young stars orbit that central binary \citep{Duchene2003,Boden2012}. The orbit of the nearest of the two has recently been modeled by \cite{Boden2012}. By combining the absolute positions of the primary provided by VLBA observations with the global orbit modeling of the system,  \cite{Torres2012} estimated the proper motion of the system's barycenter to be \mura\ = 8.3 $\pm$ 0.5 \mpy; \mudec\ = --23.6 $\pm$ 0.5 \mpy. The (heliocentric) radial velocity of the barycenter of the system was estimated by A.J.\ Boden (private communication) to be +16.38 $\pm$ 0.52 km s$^{-1}$.

\noindent {\bf T Tau} is an extremely well-studied triple system, located at 147.6 $\pm$ 0.6 pc \citep{Loinard2007}. The optically visible and classical T Tauri star is orbited by an infrared companion \citep[e.g.][]{Duchene2002} called T Tau S (the``original'' T Tauri star has now been renamed T Tau N). T Tau S is itself a tight binary system, where complex orbital motions have been detected both in the infrared and the radio \citep{Schaefer2014,Loinard2003, Loinard2007}. Since the separation between T Tau S and T Tau N is large (0.7 arcsec, corresponding to about 100 AU), the proper motion of T Tau N can reasonably be used as a proxy for the proper motion of the entire system. We adopt the value measured by \cite{Loinard2003}: \mura\ = 12.2 $\pm$ 0.6 \mpy; \mudec\ = --12.7 $\pm$ 0.6 \mpy. The (heliocentric) radial velocity of T Tau N was measured to be +19.2 $\pm$ 0.4 km s$^{-1}$ by \cite{NGuyen2012}. This is highly compatible with the older value of +19.1$\pm$1.2 km s$^{-1}$ reported by \cite{Hartmann1986}.

\noindent {\bf L1551 IRS5} is a protostellar binary located in the eponymous dark cloud Lynds 1551 to the south-east of the Taurus complex. Given its proximity to T Tau, we will adopt a similar distance for Lynds 1551, albeit with an increased uncertainty: 147 $\pm$ 5 pc. L1551 IRS5 is composed of two protostars separated by about 0.3 arcseconds (about 45 AU), presumably in relative orbit. However, given the fairly large separation between the two protostars, the orbital motions are small. The absolute and relative astrometry of these sources has been studied with the VLA by \cite{Rodriguez2003}. We will adopt the average of the two proper motions as the proper motion for the system as a whole: \mura\ = 13.2 $\pm$ 1.6 \mpy; \mudec\ = --21.2 $\pm$ 2.5 \mpy. \cite{Fridlund2002} report on high spectral resolution observations of the circumbinary disk surrounding the VLA sources, from which a (LSR) value of +6.3 $\pm$ 1.0 km s$^{-1}$ can be estimated for the systemic radial velocity. This corresponds to +18.3 $\pm$ 1.0 km s$^{-1}$ in the heliocentric frame.

\textbf{DG Tau and DG Tau B} are located near one another (they are separated by less than 1 arcmin), but they do not form a bound system. On the plane of the sky, they are located about mid-way between L1495 and HP Tau, so we will follow \cite{Rodriguez2012a} in 
adopting a distance of 150 $\pm$ 5 pc (intermediate between 130 pc for L1495 and 160 pc for HP Tau). While DG Tau is a K6 classical T Tauri star, DG Tau B is a somewhat younger Class I protostar \citep{Watson2004, Luhman2010}. The proper motion of DG Tau measured with the VLA is \mura\ = 7.5 $\pm$ 0.9 \mpy; \mudec\ = --19.0 $\pm$ 0.9 \mpy\ \citep{Rodriguez2012a}. That of DG Tau B, on the 
other hand, is \mura\ = 3.8 $\pm$ 1.9 \mpy; \mudec\ = --20.6 $\pm$ 3.3 \mpy\ \citep{Rodriguez2012b}. These are very consistent with one another, and we will adopt their weighted mean for the proper motion of the DG Tau region: \mura\ = 6.8 $\pm$ 0.8 \mpy; \mudec\ = --19.1 $\pm$ 0.9 \mpy.

The (heliocentric) radial velocity of DG Tau was measured to be +15.4 $\pm$ 1.5 km s$^{-1}$ by \cite{NGuyen2012} and $\simeq$ 16.5 km s$^{-1}$ by \cite{Bacciotti2002}. For DG Tau B, on the other hand, \cite{Zapata2015} find an LSR systemic velocity of +6.5 $\pm$ 1.0 km s$^{-1}$. This corresponds to +16.3 $\pm$ 1.0 km s$^{-1}$ in the heliocentric system. These different measurements are highly consistent with each other, and we will adopt 16.1 $\pm$ 1.0 km s$^{-1}$ for the (heliocentric) radial velocity of the DG Tau region.

\subsection{Ophiuchus}

Accurate radio proper motions are available for four sources in Ophiuchus: IRAS~16293--2422, YLW~15, S1, and DoAr21. The latter two of these sources have measured VLBI parallaxes corresponding to a distance of 120 $\pm$ 4 pc (Loinard et al.\ 2008). We will adopt this distance for all 4 sources, briefly discussing the specific case of IRAS~16293--2422 in its dedicated section. 

\noindent {\bf IRAS~16293--2422} is a multiple Class~0 protostellar system located in the dark cloud Lynds 1689N. An estimate of the distance to IRAS~16293--2422 was provided by \cite{Imai2007} who used multi-epoch VLBI observations of water masers to obtain a direct measurement of the trigonometric parallax. They obtain $\varpi$ = 5.6$^{+1.5}_{-0.5}$ mas, corresponding to $d$ = 178$^{+18}_{-37}$ pc. However, more recent VLBA water measurements by S.\ Dzib (private communication) are consistent with a shorter distance of order 120 pc, which is also the distance estimated by \cite{Loinard2008} for the Ophiuchus core. It is important to mention that water masers in low-mass star forming regions are weak, highly variable, and with short active phases \citep[e.g.,][]{Claussen1996, Desmurs2009}. As a consequence, parallaxes obtained using water masers in low-mass star-forming regions are less reliable than those measured from continuum observations of magneticaly active stars (such as S1 and DoAr21 as repoted by Loinard et al. 2008). Thus, we will adopt 120 pc for the distance to IRAS~16293--2422.

Both the absolute and the relative proper motions of the three protostars in the IRAS~16293--2422 system have been measured using multi-epoch VLA observations by \cite{Chandler2005}. Two of these protostars (A2 and B) share similar absolute proper motions, while the proper motion of the third object (A1) is significantly different. Following \cite{Loinard2002} and \cite{Chandler2005}, we adopt the mean proper motion of A2 and B for the proper motion of the system as a whole, and ascribe the different value measured for A1 to a significant contribution from its orbital motion. Thus, the proper motion adopted for IRAS~16293--2422 is \mura\ = --16.2 $\pm$ 0.9 \mpy; \mudec\ = --7.0 $\pm$ 1.1 \mpy\ \citep{Chandler2005}. 

In interferometric millimeter wavelength observations, IRAS~16293--2422 is resolved into two cores: one containing the B protostar, and the other containing the A1 and A2 objects \citep{Mundy1992}. These two condensations have slightly different radial velocities \citep{Jorgensen2011}. Since protostar B is known from the proper motion measurements to move little relative to the center of mass of the system, we will adopt the radial velocity of component B as a proxy for that of the entire system. This corresponds to $V_{lsr}$ = $+$2.7 $\pm$ 1.9 km s$^{-1}$, and is equivalent to $V_r$ = --7.7 $\pm$ 1.9 km s$^{-1}$ in the heliocentric system. 

\noindent {\bf YLW 15} (IRAS 16244--2434, IRS 43) is a binary Class~I protostar \citep{Andre1993} located in the dark cloud L1682B, near the Ophiuchus core.  The proper motion of both members of YLW 15 have been measured from VLA observations by \cite{Curiel2003} who showed further that source VLA1 is the primary of the system, whereas VLA2 is a lower-mass companion. Thus, we will adopt the proper motion of VLA1 as a proxy for the proper motion of the entire system: \mura\ = --1.4 $\pm$ 0.5 \mpy; \mudec\ = --20.8 $\pm$ 0.6 \mpy

In the DCO$^+$ maps of Ophiuchus presented by \cite{Loren1990}, YLW15 is embedded in the molecular clump F, whose radial velocity is reported as $V_{lsr}$ = $+$3.7 $\pm$ 0.7 km s$^{-1}$. We will adopt this value of YLW 15 itself, which corresponds to --6.5 $\pm$ 0.7 km s$^{-1}$ in the heliocentric reference frame. The assumption that the radial velocity of the molecular gas surrounding the star can be taken as a proxy of the radial velocity of the star itself is supported by the results of Loinard et al.\ (2008) and Torres et al.\ (2009, 2012).

\noindent {\bf S1} (IRAS 16235-2416, ROX 14, YLW 36) is located in the Ophiuchus core (Lynds 1688). It is a B4 star with a mass of about 6 M$_\odot$, and it is among the brightest red, near-infrared, far-infrared, X-ray and radio sources in the region \citep{Grasdalen1973, Fazio1976, Montmerle1983, Leous1991, Loinard2008}. Its proper motion has been measured using multi-epoch VLBA observations by \cite{Loinard2008}: \mura\ = --3.88 $\pm$ 0.87 \mpy; \mudec\ = --31.55 $\pm$ 0.69 \mpy. 

We did not find any direct (photospheric) radial velocity measurement for S1 in the literature. However, in the DCO$^+$ observations reported by \cite{Loren1990}, S1 is located on the edge of clump  A. The mean radial velocity of the DCO$^+$ emission for clump A is 3.5 km s$^{-1}$ measured in the LSR. The mean width of the DCO$^+$ lines, on the other hand, is 1.0 km s$^{-1}$, so we adopt $v_{lsr}$ = 3.5 $\pm$ 1.0 km s$^{-1}$ for the radial velocity of this source. This corresponds to $V_r$= --6.7 $\pm$ 1.0 km s$^{-1}$ in the heliocentric reference frame.

\noindent {\bf DoAr 21} (V2246 Oph, Haro 1-6, HBC 637, ROX 8, YLW 26) also belongs to the Ophiuchus core. It is a $\sim$ 2.2 M$_\odot$ star of spectral type K1 with an infrared excess around 25 $\mu m$ attributed to a circumstellar disk \citep{Jensen2009}. It is associated with a strongly variable radio source and a bright X-ray source \citep{Montmerle1983, FeigelsonMontmerle1985, Dzib2013}. The proper motion has been measured by \cite{Loinard2008}: \mura\ = --26.47 $\pm$ 0.92 \mpy; \mudec\ = --28.23 $\pm$ 0.73 \mpy. The radial velocity derived from optical spectroscopy was provided by \cite{Massarotti2005}: $V_r$ = -4.6 $\pm$ 3.3 km s$^{-1}$ (heliocentric). This is consistent with the value --6 $\pm$ 4 km s$^{-1}$ reported by \cite{Jensen2009}.

\bigskip

A summary of the proper motion and radial velocity measurements detailed above is provided as Table 1. For completeness, we also include the proper motions converted to Galactic $(\ell,b)$ coordinates. From that summary, it is clear that the radial velocities are typically accurate to about 1 km s$^{-1}$. The proper motions, on the other hand, typically have a 1-dimensional uncertainty of 1 mas yr$^{-1}$. At the distance of Ophiuchus and Taurus, this also corresponds to about 1 km s$^{-1}$ errors on the tangential velocity. 

\section{Analysis}

\subsection{Determination of the 3D velocity vectors}

Since our goal here is to analyze the internal and relative motions of Taurus and Ophiuchus, we first convert the measured proper motions and radial velocities to 3-dimensional velocity vectors. We will express these vectors in the rectangular $(x,y,z)$ coordinate system commonly used for Galactic studies. The origin of the system is at the Sun; the $(Ox)$ axis runs along the Sun -- Galactic center direction, the positive direction being toward the Galactic center; $(Oy)$ is in the Galactic plane, orthogonal to $(Ox)$, with the positive direction in the direction of Galactic rotation; $(Oz)$ is perpendicular to the Galactic plane, oriented toward the Galactic North Pole, thereby making $(Oxyz)$ a right-handed coordinate system. From the data in Table 1, both the positions $(X,Y,Z)$ and the heliocentric velocities $(U,V,W)$ of each of our stars in the $(x,y,z)$ frame can easily be computed. They are listed in Table 2.

Expressing velocities in the heliocentric system is practical from an observational point of view, because the dynamics of the Solar System are so well known that heliocentric velocities only contain extremely small systematic uncertainty (i.e.\ the transformation from topocentric to heliocentric velocities is very accurate). For the point of view of Galactic Dynamics, however, a Sun-based system is not ideal. In particular, for objects in the Solar neighborhood, the Local Standard of Rest (LSR) is preferable. The transformation from heliocentric to LSR velocities is effected by subtracting the motion of the Sun relative to the LSR from the heliocentric velocities. This is straightforward in principle, but introduces significant errors in practice, because the Solar motion relative to the LSR is somewhat uncertain. As we mentioned earlier, velocities reported in the LSR system normally assume a Solar motion of +20 km s$^{-1}$ toward B1900 equatorial coordinates $(18^h; +30^\circ)$. More recent determinations, however, suggest significantly different values. Until recently, the Hipparcos-based determination of \cite{DehnenBinney1998} was widely used. For this determination, the components of the Solar motion in the rectangular $(x,y,z)$ coordinate system introduced earlier are $U_0$ = 10.00 $\pm$ 0.36 km s$^{-1}$, $V_0$ = 5.25 $\pm$ 0.62 km s$^{-1}$, and $W_0$ = 7.17 $\pm$ 0.38 km s$^{-1}$. On the basis of a global analysis of high accuracy trigonometric parallaxes to high-mass star-forming regions distributed across the Galactic plane, \cite{Reid2009} argued in favor of a significantly larger value of $V_0$. This is supported by a recent re-analysis of stellar kinematics in the Solar neighborhood by \cite{Schonrich2010} who obtained $U_0$ = 11.1 $\pm$ 0.7 km s$^{-1}$, $V_0$ = 12.2 $\pm$ 0.47 km s$^{-1}$, and $W_0$ = 7.25 $\pm$ 0.37 km s$^{-1}$. Here, we will adopt this latter value to transform the $(U,V,W)$ heliocentric velocities of the young stars in Taurus and Ophiuchus into $(u,v,w)$ LSR velocities for those stars. The results are given in Table 2.

\subsection{Taurus internal kinematics}

Let us now analyze the 3D velocity vectors in the Taurus complex. In the top row of Figure 1, we show their projections onto the $(Oxy)$, $(Oxz)$, and $(Oyz)$ planes. When we consider the heliocentric velocities (shown as green arrows in the top row of Figure 1), the motions appear highly organized as a result of the dominant reflex motion induced by the Sun. When the Solar motion is removed (magenta arrows in the top row of Figure 1), the motions appear less clearly organized, although there is still a clear remaining bulk motion, particularly in the negative $(Ox)$ direction. This bulk motion is shown as a blue arrow in the top row of Figure 1, and will be discussed further below. The smaller value and more disorganized aspect of the LSR velocities compared with the heliocentric ones evidently reflects the fact that the heliocentric velocities are dominated by (minus) the Solar motion itself. While the Sun has a 15--20 km s$^{-1}$ non-circular (i.e.\ peculiar) velocity component in its orbit around the Galactic center, the Taurus complex is on a much more circular orbit, as expected for a region of star-formation.

To characterize the internal kinematics of the stars in the Taurus complex, we now compute the difference $(\delta u,\delta v,\delta w)$ between the velocity $(u,v,w)$ of each star in Taurus and their mean $(\overline{u},\overline{v},\overline{w})$. For reference, the latter is $(\bar{u},\bar{v},\bar{w})$ = $(-5.6,-0.6,-2.0)$ km s$^{-1}$, and corresponds to $(\overline{U},\overline{V},\overline{W})$ = $(-16.7,-12.9,-9.2)$ km s$^{-1}$ when expressed in heliocentric velocities. This is very similar to the value obtained independently by \cite{BertoutGenova2006} from a larger sample of young stars in Taurus with lower accuracy proper motion and distance measurements: $(\overline{U},\overline{V},\overline{W})$ = $(-15.4,-11.7,-9.9)$ km s$^{-1}$. The projections of $(\delta u,\delta v,\delta w)$ are shown in the bottom row of Figure 1. They have a fairly random appearance, with one-dimensional dispersion $\sigma_u$ = 1.3 km s$^{-1}$, $\sigma_v$ = 2.1 km s$^{-1}$, $\sigma_w$ = 3.2 km s$^{-1}$. The three-dimensional velocity dispersion is $\sigma$ = $\sqrt{\sigma_u^2+\sigma_v^2+\sigma_w^2}$ = 4.1 km s$^{-1}$.

To assess quantitatively the relative importance of random and organized motions within Taurus, we proceed as follows. Each star is located at a position relative to the center of the complex given by the vector $\mathbf{r_*}$ and moves relative to the complex at a velocity $\mathbf{\delta v_*}$. To each position vector $\mathbf{r_*}$, we associate the unit vector $\mathbf{\hat{r}_*}$ = $\mathbf{r_*}/|\mathbf{r_*}|$ which simply points from the center to each given star in the complex. We will consider two types of organized motions: expansion (or contraction) and rotation. The velocity $\mathbf{\delta v_*}$ of each star in the complex (measured relative to the complex itself) should be parallel to $\mathbf{\hat{r}_*}$ for expansion, and anti-parallel for contraction. Thus, the dot product  $\mathbf{\hat{r}_*} ~ . ~\mathbf{\delta v_*}$ should be large and positive for expansion, and large but negative for contraction. By the same token, the cross product $\mathbf{\hat{r}_*} ~\times~ \mathbf{\delta v_*}$ should be small for expansion and contraction.

Conversely, for large-scale rotation, we expect the cross product $\mathbf{\hat{r}_*} ~\times~ \mathbf{\delta v_*}$ to be large and the dot product $\mathbf{\hat{r}_*} ~ . ~\mathbf{\delta v_*}$ to be small. For instance, for circular rotation in a disk-like structure, $\mathbf{\delta v_*}$ and $\mathbf{\hat{r}_*}$ would be orthogonal, so the dot product would be zero and the cross product would be maximum. For a 3-dimensional structure such as Taurus, the situation would be slightly more complex, but one would certainly expect the cross product to be large and the dot product to be small. An alternative way of looking at this issue is that the quantity $\mathbf{\hat{r}_*} ~\times~ \mathbf{\delta v_*}$ is a proxy for the specific angular momentum of the complex, which is expected to be large for rotation, but small for contraction and expansion.

We calculated the cross and dot products described above for each star in Taurus, and took their mean. Notice that both quantities have dimensions of velocity (this was, indeed, the reason for using the unit vector $\mathbf{\hat{r}_*}$ rather than $\mathbf{r_*}$ itself, in the dot and cross products). Because the dot product is a measure of expansion, while the cross product is a measure of rotation, we will introduce the following definitions:

\[ v_{exp} = \overline{\mathbf{\hat{r}_*} ~ . ~\mathbf{\delta v_*}}, \]

\[ v_{rot} = \overline{\mathbf{\hat{r}_*} ~ \times ~\mathbf{\delta v_*}}. \]

\noindent
Of course, these quantities are not strictly expansion and rotation velocities, but in view of our previous discussion, they can be used as proxies for them.

For Taurus, we obtain $v_{exp}$ = --0.15 km s$^{-1}$ and $v_{circ}$ = $(-1.55, +2.03, -0.02)$ km s$^{-1}$. The individual values of the dot and cross products are shown in Table 3, and the projections of the individual cross product vectors are shown in the bottom row of Figure 1. The expansion velocity appears very small compared with the velocity dispersion of $\sim$ 4 km s$^{-1}$ measured earlier. This results from the fact that the individual dot products are alternatively positive and negative (see Table 3), resulting in a small net mean. Thus, in the radial direction, the stellar motions appear to be dominated by a random component rather than by an organized expansion or contraction pattern. This is correctly reflected by the small absolute value of $v_{exp}$.

The situation for rotation is clearly different. We obtain $v_{rot}$ of about 2 km s$^{-1}$ comparable with the velocity dispersion. Moreover, the individual cross product vectors are clearly not randomly oriented. Instead, their components along the $(Ox)$ axis are systematically negative, their components along the $(Oy)$ axis are systematically positive, while their component along the $(Oz)$ axis are around zero (see Table 3, Figure 1). This suggests that the entire Taurus complex is tumbling with a rotation velocity $\omega$ in the $(Oxy)$ plane. Since Taurus is a few tens of pc across and the rotation velocity is a few km s$^{-1}$, the rotation of Taurus induces velocity gradients of order 0.1 km s$^{-1}$ pc$^{-1}$ across the complex.

The relevance of rotation to the equilibrium of the Taurus complex can be estimated by assuming a homogeneous spherical cloud with $v_{rot}$ = 2 km s$^{-1}$ at its edge and comparing the gravitational energy with the rotational energy. We assume that Taurus has a total mass of $M \simeq 3 \times 10^4~M_\odot$ \citep{Ungerechts1987} and a radius of $R \simeq$ 15 pc \citep{Guedel2007}.

The gravitational energy is given by

$$E_{grav} \simeq - {{3 G M^2} \over {5 R}} \simeq -3 \times 10^{48}~ergs,$$

while the rotational energy is

$$E_{rot} \simeq {{M v_{rot}^2} \over {5}} \simeq 5 \times 10^{47}~ergs.$$

We then conclude that rotation plays a minor role in the virial equilibrium of Taurus.

Our analysis of the internal kinematics of Taurus is based on high accuracy radial velocity, proper motion, and distance measurements of a limited sample of young stellar objects. The comparison (mentioned in Section 3.2) of the mean bulk motion of Taurus measured here with the determination by \cite{BertoutGenova2006} based on a much larger sample (but with much less accurate astrometric information) shows that our conclusions 
are trustworthy. It will be very interesting to repeat our analysis with larger samples of young stellar objects when they become available. For instance, the Gould's Belt Distances Survey \citep{Loinard2011} will provide parallaxes and proper motion measurements similar to those 
used here for tens of young stars in Taurus and other regions. The GAIA mission \citep{deBruijne2012} will provide data with similar 
accuracy at least for YSOs that are not too deeply embedded into their parental dusty cocoons. 

\subsection{The Ophiuchus bulk motion}

The previous analysis could be repeated for Ophiuchus, but the results for internal kinematics would be quite uncertain, because there are only four stars with accurate proper motions, the stars are highly concentrated (within a few pc of each other), and the necessary (but poorly justified for 2 of the 4 stars) assumption that all are at a common distance. We will defer this analysis to a forthcoming paper where additional astrometric results from the Gould's Belt Distances Survey \cite{Loinard2011} will be incorporated \citep{Ortiz-Leon2015}. Here, we will merely use the Ophiuchus results to estimate the mean bulk motion of the region. We obtain $(\bar{u},\bar{v},\bar{w})$ = $(+4.3,-0.9,+2.4)$ km s$^{-1}$, which corresponds to $(\overline{U},\overline{V},\overline{W})$ = $(-6.8,-13.1,-4.8)$ km s$^{-1}$ when expressed in heliocentric velocities. 

\subsection{The relative motion between Taurus and Ophiuchus}

Taurus and Ophiuchus are fortuitously located almost symmetrically with respect to the Sun: Taurus lies at $\sim$ 145 pc in the direction of the Galactic anti-center, at a Galactic latitude $\sim$ --15$^\circ$. Ophiuchus, on the other hand, lies at about 120 pc in the direction of the Galactic center, at a Galactic latitude $\sim$ --15$^\circ$. In the rectangular system that we use throughout this paper, the mean position of the stars that we observed in Taurus is $(-134.9, +16.8, -42.1)$ pc, while the mean position of the stars in Ophiuchus is $(+114.2,  -13.7, + 34.5)$ pc. These two positions are almost exactly opposite to one another in the rectangular frame where the Sun is at the origin. Remarkably, the mean velocity of the stars in Taurus that we calculated earlier ($v_{tau}$ = $(-5.6,-0.6,-2.0)$ km s$^{-1}$) and of those in Ophiuchus ($v_{oph}$ = $(+4.3,-0.9,+2.4)$ km s$^{-1}$) are also almost exactly opposite to one another (Figure 2). Both the angle between $v_{tau}$ and the line joining Ophiuchus to Taurus, and the angle between $v_{oph}$ and the line joining Taurus to Ophiuchus are of the order of 13$^\circ$ and consistent within the errors with 0$^\circ$. This strongly suggest a common origin for Taurus and Ophiuchus (Figure 2). Indeed, running the time backwards, we find that Taurus and Ophiuchus must have been very near each other about 23.7 Myr ago (this is assuming a constant velocity).

Most young stars and molecular cloud complexes in the Solar neighborhood are distributed  within an expanding structure inclined by about 15--20$^\circ$ from the Galactic plane, and known as Gould's Belt (see \cite{Poppel1997} for an extensive review). The putative center of this structure is located in the Galactic mid-plane, about 100 pc from the Sun in the direction of the Galactic anti-center \citep{PerrotGrenier2003}. With this assumed center, all nearby substantial star-forming regions in  the Solar neighborhood except Taurus can be accommodated on an elliptical ring inclined by 17.2$^\circ$ from the Galactic plane, with semi-major and minor axes of 373 and 233 pc, respectively, and a line of node at $\ell$ = 296$^\circ$. \citep{PerrotGrenier2003}. The corresponding dynamical age of the structure is 26.4 Myr, remarkably similar to the dynamical age that we derived above for the Taurus-Ophiuchus system. Yet, the relation between Taurus and Gould's Belt is somewhat unclear. Taurus appears to be {\em projected} in the direction of Gould's Belt (and is indeed often included in Gould's Belt surveys), at a location intermediate between Perseus and Orion. However, it is not contained in the ring that defines Gould's Belt. Instead, it is located near the center of the Belt \citep[see e.g.\ Figure 5 in][]{PerrotGrenier2003}. 

A possible explanation for the peculiar location of Taurus with respect to Gould's Belt was proposed by \cite{OlanoPoppel1987} ; see also the review by \cite{Poppel1997}. In that scheme, the Taurus material would have been ejected from a region located somewhere along the ring containing the star-forming regions in Gould's Belt. They argue in favor of a region at $\ell$ $\sim$ 245$^\circ$ and $b$ $\sim$ -14$^\circ$. Our analysis of the relative kinematics between Ophiuchus and Taurus would be inconsistent with this original position (which would instead have to lie fairly close to the current position of the Sun), but the mechanism proposed by \cite{OlanoPoppel1987} could still provide the correct theoretical framework for the observations. In this scheme, Taurus and Ophiuchus would originate as the result of an energetic event which would have occurred roughly simultaneously with (or only a few Myr after) the creation of Gould's Belt and which would have launched interstellar material on opposite ballistic trajectories. It is interesting in this respect to consider the energetics. In combination, Taurus and Ophiuchus contain about 5 $\times$ 10$^4$ $M_\odot$ in material, and they are both moving at about 5 km s$^{-1}$. This corresponds to a total kinetic energy of 2 $\times$ 10$^{49}$ ergs, which is only a fraction of the kinetic energy output of a typical core-collapse Supernova explosion. 

\section{Conclusions}

In this paper, we have combined radial velocity measurements with high accuracy proper motion and parallax determinations for a sample of young stars in Taurus and Ophiuchus to characterize both their internal kinematics and their relative motion. We find no evidence for contraction or expansion of the Taurus complex but fairly conclusive indications for global rotation in Taurus. These conclusions will be strengthened once additional high quality parallaxes and proper motions become available for young stars in Taurus and Ophiuchus first as part of the Gould's Belt Distances Survey \citep{Loinard2011} and then from the GAIA mission.

In addition, we measure the relative velocity of Taurus and Ophiuchus and show that they are moving away from each other at a velocity of order 5 km s$^{-1}$. This points to a common origin, some 23.7 Myr ago, possibly related to the phenomena that gave birth to Gould's Belt.

\acknowledgments
J.L.R., L.L., S.D., G.N.O, and L.F.R. acknowledge the financial support of DGAPA, UNAM, and CONACyT, Mexico. The National Radio Astronomy Observatory is operated by Associated Universities Inc. under cooperative agreement with the National Science Foundation. 

{}

\clearpage

\begin{deluxetable}{lcccccc}
\tabletypesize{\scriptsize}
\tablecaption{Observational data\label{propermotion}}
\tablewidth{0pt}
\tablehead{
\colhead{Source} & \colhead{$d$ (pc)} & \colhead{\mura\ (\mpy)} & \colhead{\mudec\ (\mpy)} & \colhead{$\mu_{\ell} cos{b}$ (\mpy)}&
 \colhead{$\mu_{b} $ (\mpy)} & \colhead{$V_r$ (km s$^{-1}$)}
}
\startdata
IRAS 16293-2422 & 120.0 $\pm$ 4.0 & -16.2 $\pm$ 0.9 & -7.0 $\pm$ 1.1 &-15.85$\pm$1.06&7.77$\pm$1.1& -7.7 $\pm$ 1.9 \\%
YLW 15 & 120.0 $\pm$ 4.0 & -1.4$\pm$ 0.5 & -20.8 $\pm$ 0.8 &-16.55$\pm$0.58&-12.68$\pm$0.58& -6.5 $\pm$ 0.7 \\%
S1 & 120.0 $\pm$ 4.0 & -3.88 $\pm$ 0.87 & -31.55 $\pm$ 0.69 &-26.25$\pm$0.81&-17.95$\pm$0.78& -6.7 $\pm$ 1.0 \\%
DoAr 21 & 120.0 $\pm$ 4.0 & -26.47 $\pm$ 0.92 & -28.23 $\pm$ 0.73 &-38.69$\pm$0.86&0.85$\pm$0.84& -4.6 $\pm$ 3.3 \\%
Hubble 4 & 132.8 $\pm$ 0.5 & +4.30 $\pm$ 0.05 & -28.9 $\pm$ 0.3 &23.95$\pm$0.25&-16.75$\pm$0.24& +18.0 $\pm$ 2.0 \\%
HDE~283572 & 128.5 $\pm$ 0.6 & +8.88 $\pm$ 0.06 & -26.6 $\pm$ 0.1 &25.53$\pm$0.09&-11.61$\pm$0.09& +14.2 $\pm$ 1.0 \\%
HP Tau G2 & 161.2 $\pm$ 0.9 & +13.90 $\pm$ 0.06 & -15.6 $\pm$ 0.3 &20.89$\pm$0.25&0.74$\pm$0.21& +16.6 $\pm$ 1.7 \\%
V 773 Tau & 132.8 $\pm$ 2.3 & +8.3 $\pm$ 0.5 & -23.6 $\pm$ 0.5 &22.72$\pm$0.51&-10.48$\pm$0.53& +16.32 $\pm$ 0.52 \\%
T Tau N & 146.7 $\pm$ 0.6 & +12.2 $\pm$ 0.6 & -12.7 $\pm$ 0.6 &17.59$\pm$0.64&0.95$\pm$0.63& +19.2 $\pm$ 0.4 \\%
L1551 IRS5 & 147.0 $\pm$ 5.0 & +13.2 $\pm$ 1.6 & -21.2 $\pm$ 2.5 &24.78$\pm$2.18&-3.12$\pm$1.91& +18.3 $\pm$ 1.0 \\%
DG Tau (A+B) & 150.0 $\pm$ 5.0 & +6.8 $\pm$ 0.8 & -19.1 $\pm$ 0.9 &18.79$\pm$0.91&-7.62$\pm$0.91& +16.1 $\pm$ 1.0 \\%
\enddata
\tablecomments{The proper motions in columns 3 and 4 are expressed in equatorial $(\alpha,\delta)$ coordinates, while those in columns 5 and 6 are in Galactic $(\ell,b)$ coordinates.}
\end{deluxetable}

\clearpage

\begin{deluxetable}{lccccccccc}
\tabletypesize{\scriptsize}
\tablecaption{Derived velocities\label{positions}}
\tablewidth{0pt}
\tablehead{
\colhead{Source} & \colhead{$U$} & \colhead{$V$} & \colhead{$W$} & \colhead{$u$}& \colhead{$v$}&\colhead{$w$}&
\colhead{$X$}& \colhead{$Y$}&\colhead{$Z$}\\
\colhead{} & \colhead{} & \colhead{\kmps} & \colhead{} & \colhead{}& \colhead{\kmps}&\colhead{}&
\colhead{}& \colhead{pc}&\colhead{}
}
\startdata
IRAS~16293--2422 &-9.52&-8.06&2.15&1.59&4.19&9.4&114.85&-12.21&32.77\\%
YLW 15 &-5.3&-8.84&-8.77&5.81&3.41&-1.52&114.22&-14.01&34.21\\%
S1 &-5.22&-14.42&-11.72&5.89&-2.18&-4.47&114.04&-13.81&34.89\\%
DoAr 21 &-7.36&-21.25&-0.93&3.75&-9.01&6.33&113.53&-14.87&36.09\\%
Hubble 4 &-17.17&-11.98&-14.98&-6.07&0.27&-7.73&-125.54&24.77&-35.54\\%
HDE~283572 &-14.55&-13.1&-10.52&-3.45&-0.86&-3.27&-122.01&22.91&-33.34\\%
HP Tau G2 &-17.24&-14.73&-4.11&-6.14&-2.49&3.15&-154.43&11.54&-45.1\\%
V 773 Tau &-16.5&-11.17&-10.94&-5.4&1.08&-3.69&-124.76&26.03&-37.37\\%
T Tau N &-18.94&-11.01&-6.24&-7.84&1.24&1.02&-136.7&9.01&-52.28\\%
L1551 IRS5 &-16.77&-16.97&-8.32&-5.67&-4.73&-1.07&-138.13&2.59&-50.42\\%
DG Tau (A+B) &-15.79&-11.23&-9.57&-4.69&1.02&-2.32&-142.89&20.48&-40.52\\%
\enddata
\end{deluxetable}

\clearpage

\begin{deluxetable}{lcccc}
\tabletypesize{\scriptsize}
\tablecaption{Dot and cross products for the sources in Taurus\label{productos}}
\tablewidth{0pt}
\tablehead{
\colhead{Source} & \colhead{$\hat{r}\cdot\vec{v}$} & \colhead{} & \colhead{$\hat{r}\times\vec{v}$} & \colhead{}
}
\startdata
Hubble 4 &0.23&-1.59&4.99&1.63\\%
HDE~283572 &-2.49&-3.72&3.65&0.87\\%
HP Tau G2 &0.92&-0.36&2.12&-0.96\\%
V773 Tau &-3.19&-0.32&2.18&-1.6\\%
T Tau N &0.69&-1.65&1.27&1.06\\%
L1551 IRS5 &3.01&-2.82&0.21&0.73\\%
DG Tau (A+B) &-0.19&-0.43&-0.14&-1.86\\%
\enddata
\end{deluxetable}

\begin{figure}
\epsscale{1.0}
\plotone{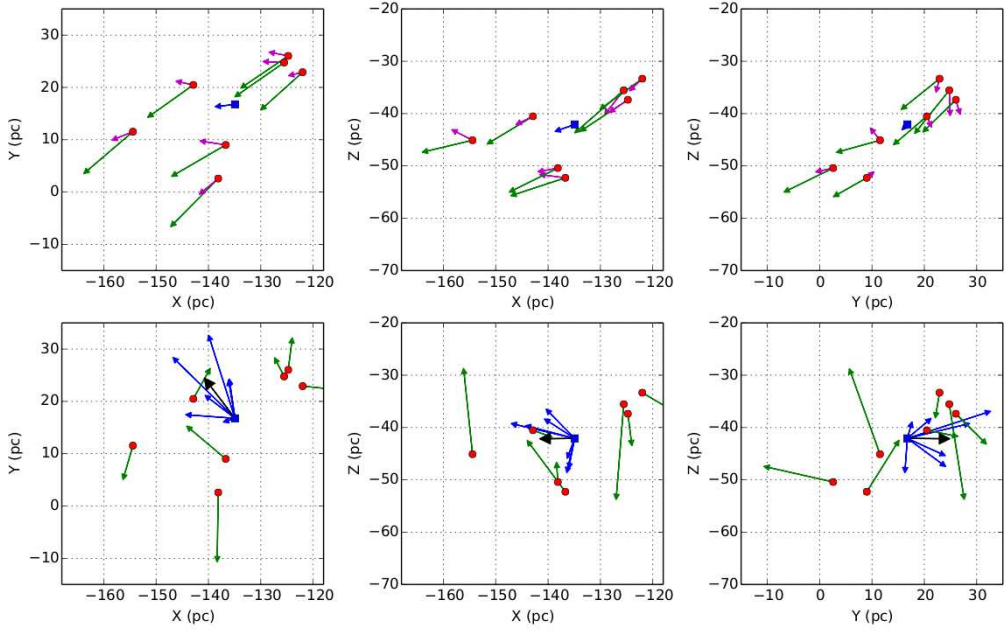}
\caption{Top: heliocentric (green) and LSR (magenta) velocities for the sources in Taurus expressed in the cartesian coordinate system described in the text. The blue arrow shows the mean LSR velocity of the Taurus complex. Bottom: The green arrows show $\delta v$, the difference between the velocity of each star and the mean velocity of the Taurus complex. The blue arrows show the $\mathbf{\hat{r}_*} ~ \times ~\mathbf{\delta v_*}$ cross product; the black arrow is the mean of these cross products.}
\label{fig2}
\end{figure}

\begin{figure}
\epsscale{1.0}
\plotone{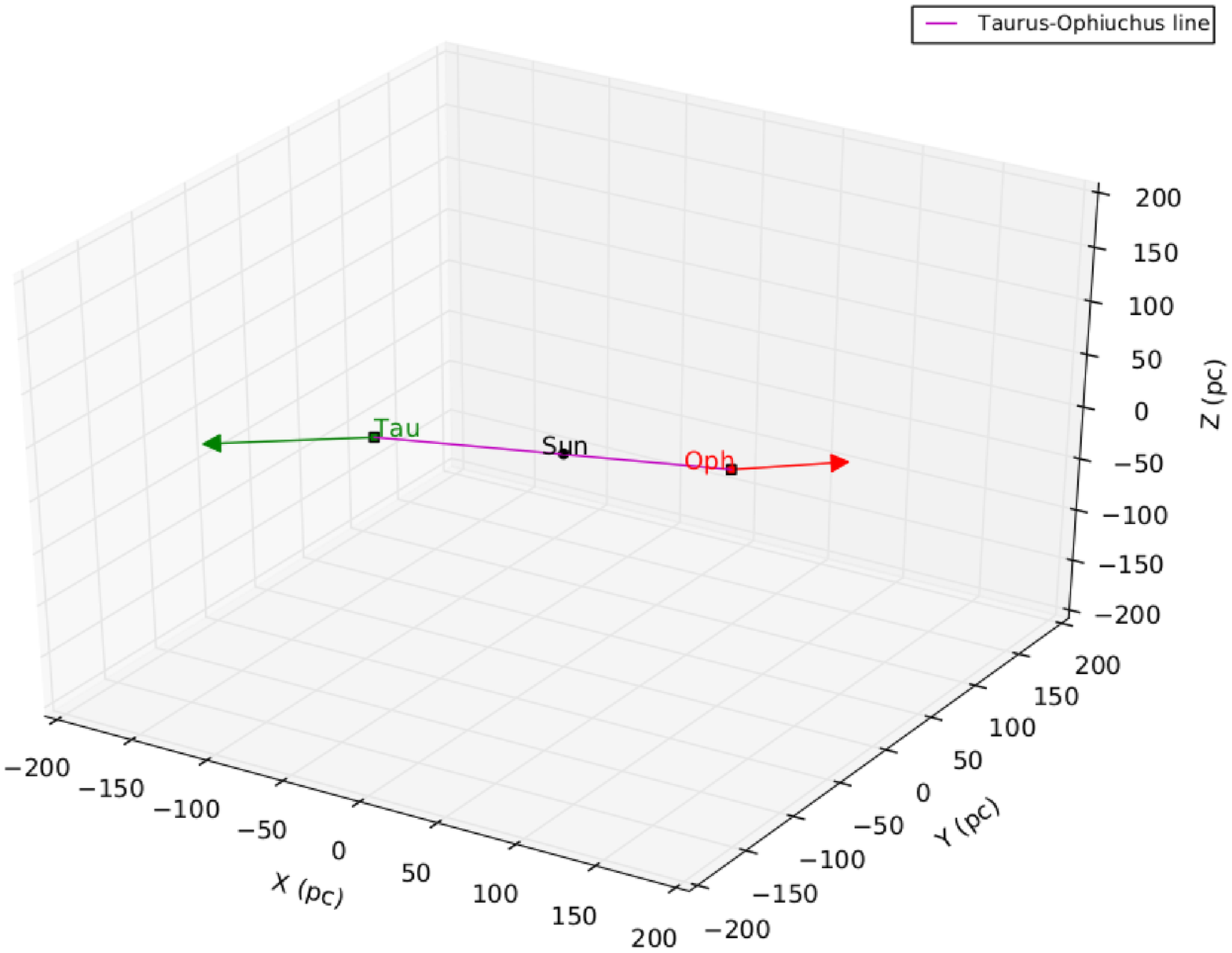}
\caption{3D rendering of the relative positions of the Sun, Taurus, and Ophiuchus; the arrows show the LSR bulk velocities of Taurus and Ophiuchus.}
\label{fig1}
\end{figure}

\end{document}